\documentclass[journal,onecolumn,12pt]{IEEEtran}
\ifCLASSINFOpdf
\else
\fi
\usepackage{amsmath}
\usepackage[dvips]{graphicx}
\usepackage[usenames,dvipsnames]{xcolor}
\usepackage{colortbl}
\usepackage{multirow}
\usepackage{pgf,tikz}
\usepackage{tkz-graph}
\usepackage{subfig}
\usepackage{tikz}
\usetikzlibrary{arrows}
\usepackage{amsmath, amsthm, amscd, amsfonts, amssymb, color}
\usepackage[bookmarksnumbered, plainpages]{hyperref}
\usepackage{booktabs,chemformula}
\usetikzlibrary{calc,patterns,angles,quotes}
\usetikzlibrary{angles,quotes,3d}
\usetikzlibrary{angles,quotes}
\usepackage{tabularx}
\usepackage{lipsum}
\usepackage{relsize}
\usepackage{multicol}
\usepackage{longtable}
\usepackage{multicol}
\usepackage{rotating}
\usepackage[labelfont=bf]{caption} 
\usepackage{subfig}
\DeclareMathOperator{\tr}{tr}
\newtheorem{thm}{Theorem}[section]
\newtheorem{cor}[thm]{Corollary}

\newtheorem{lem}[thm]{Lemma}
\newtheorem{prop}[thm]{Proposition}
\newtheorem{defn}[thm]{Definition}
\newtheorem{rem}[thm]{Remark}

\newtheorem{ex}[thm]{Example}
\numberwithin{equation}{section}

\begin{document}
%
\title{Improved  bounds on the size of  permutation codes under Kendall $\tau$-metric}
%
%
%

\author{ Farzad~Parvaresh,  
Reza~Sobhani, 
Alireza~Abdollahi, 
        Javad~Bagherian,  
        Fatemeh~Jafari 
                  and~ Maryam~Khatami 
                 
 \thanks{The work has been partially presented in IEEE Workshop on Communication and Information Theory (IWCIT) 2022 \cite{ABJKPSnew}.}                 
\thanks{F. Parvaresh is with Department of Electrical Engineering University of Isfahan, Isfahan, Iran} 
\thanks{R. Sobhani is with Department of Applied Mathematics and Computer Science, Faculty of Mathematics and Statistics, University of Isfahan, Isfahan 81746-73441, Iran}    
\thanks{A. Abdollahi, J. Bagherian, F. Jafari and M. Khatami   are with Department of Pure Mathematics, Faculty of Mathematics and Statistics, University of Isfahan, Isfahan 81746-73441, Iran}
\thanks{A. Abdollahi and F. Parvaresh are also with School of Mathematics, Institute for Research in Fundamental Sciences (IPM), 19395-5746 Tehran, Iran.} 
\thanks{E-mail addresses:   \href{mailto: a.abdollahi@math.ui.ac.ir}{a.abdollahi@math.ui.ac.ir} (A. Abdollahi),  \href{mailto: bagherian@sci.ui.ac.ir}{bagherian@sci.ui.ac.ir} (J. Bagherian), 
\href{mailto: math_fateme@yahoo.com}{math$\_$fateme@yahoo.com} (F. Jafari), \href{mailto: m.khatami@sci.ui.ac.ir}{m.khatami@sci.ui.ac.ir} (M. Khatami), \href{mailto: f.parvaresh@eng.ui.ac.ir}{f.parvaresh@eng.ui.ac.ir} (F. Parvaresh),  \href{mailto: r.sobhani@sci.ui.ac.ir}{r.sobhani@sci.ui.ac.ir} (R. Sobhani).}}
\maketitle

\begin{abstract}
In order to overcome the challenges caused by flash memories and also to protect against errors related to reading information stored in DNA molecules in the shotgun sequencing method, the rank modulation is proposed.  In the  rank modulation framework,   codewords are permutations. In this paper, we study the largest size $P(n, d)$ of permutation codes of length $n$, i.e., subsets of the set $S_n$ of all permutations on $\{1,..., n\}$ with the minimum distance at least $d\in\{1,\ldots ,\binom{n}{2}\}$ under the Kendall $\tau$-metric. By presenting an algorithm and some theorems, we managed to  improve the known lower and upper  bounds for $P(n,d)$.
 In particular, we show that $P(n,d)=4$ for all $n\geq 6$ and  $\frac{3}{5}\binom{n}{2}< d \leq \frac{2}{3} \binom{n}{2}$.  Additionally, we prove that for any prime number $n$ and integer  $r\leq \frac{n}{6}$, 
$
P(n,3)\leq (n-1)!-\dfrac{n-6r}{\sqrt{n^2-8rn+20r^2}}\sqrt{\dfrac{(n-1)!}{n(n-r)!}}.
$
This result greatly improves  the upper bound of $P(n,3)$ for all primes  $n\geq 37$.
\end{abstract}


%
\IEEEpeerreviewmaketitle

%
%
%
\section{Introduction}
In order to overcome the challenges caused by flash memories and also to protect against errors related to reading information stored in DNA molecules in the shotgun sequencing method, the rank modulation is proposed (see  \cite{jiang} and \cite{kiah}, respectively). In the  rank modulation framework,   codewords are permutations.
 Within this framework, permutation codes were extensively examined using three metrics: the Kendall $\tau$-metric \cite{ABJKPS,jiang,WZYG,WWYF,V}, the  Ulam metric \cite{FA} and the  $\ell_\infty  $ metric \cite{KL,WF}. This study specifically concentrates on permutation codes under the Kendall $\tau$-metric.

A Permutation Code (PC) of length $n$ represents a non-empty subset of $S_n$, which includes all permutations of the set $[n] := \{1, 2, \ldots, n\}$.
In the context of a permutation $\pi:=[\pi(1),\pi(2),\ldots,\pi(i),\pi(i+1),\ldots,\pi(n)]\in S_n$, an adjacent transposition, denoted as $(i, i + 1)$ for $1\leq i\leq n-1$, transforms $\pi$ into the permutation $[\pi(1),\pi(2),\ldots,\pi(i+1),\pi(i),\ldots,\pi(n)]$.
The Kendall $\tau$-distance between two permutations, $\rho$ and $\pi$ in $S_n$, is defined as the minimum number of adjacent transpositions required to express $\rho \pi^{-1}$ as their product.
In the context of the Kendall $\tau$-metric, a PC of length $n$ with  minimum distance  $d$ can  correct up to $\frac{d-1}{2}$ errors induced by charge-constrained errors, as cited in \cite{jiang}.

A central question in the theory of PCs is  determining  the value of $P(n,d)$, that is the size of the largest code in $S_n$ with minimum Kendall $\tau$-distance $d$, for $d\leq \binom{n}{2}$. The exact value of $P(n,d)$ is determined for $d\in\{1,2\}$ and $\frac{2}{3}\binom{n}{2}< d\leq \binom{n}{2}$ \cite{BE} and also for $n=5$  and for $n=6$ when $d\neq 3$ \cite{V}. Furthermore, several researchers have presented bounds on $P(n, d)$ (see \cite{ABJKPS,barg,BE,jiang,WZYG,WWYF,V}). \\
In this paper,    we   present a theorem supporting the  value of $P(n,d)$ as follows:
\begin{thm}\label{size4}
 $P(n,d)= 4$, for all $n\geq 6$ and  $\frac{3}{5}\binom{n}{2}< d \leq \frac{2}{3} \binom{n}{2}$. 
\end{thm}
Moreover, we achieved significant improvements on the lower bound of $P(n,d)$ when $n\in\{7,8\}$ by constructing new PCs from the subgroups of $S_n$ (see Table \ref{t1}, below) and, in particular,  we establish  $P(7,12)=7$.\\
Utilizing sphere packing bound (see \cite[Theorems 12 and 13]{jiang}),   $P(n,3)\leq (n-1)!$. In \cite[Corollary 2.5  and Theorem 2.6]{white} and \cite[Corollary 2]{BE}, it is proved that if $n>4$ is a prime number or $4\leq n\leq 10$, then $P(n,3)\leq (n-1)!-1$. Enhancing this, in \cite[Theorem 1.1]{ABJKPS}, we improved the upper bound to   $P(n,3)\leq (n-1)!-\lceil \frac{n}{3}\rceil+2\leq (n-1)!-2$ for all primes $n\geq 11$. Here we prove an additional upper bound on $P(n,3)$ as follows:
\begin{thm}\label{bound}
For  a prime number $n$ and  integer $r\leq \frac{n}{6}$, 
\begin{equation}\label{mainrelation}
P(n,3)\leq (n-1)!-\dfrac{n-6r}{\sqrt{n^2-8rn+20r^2}}\sqrt{\dfrac{(n-1)!}{n(n-r)!}}.
\end{equation}
\end{thm}
The upper bound  for $P(n,d)$ derived from \cite[Theorem 1.1]{ABJKPS} outperforms that from Theorem \ref{bound} for all prime numbers $11\leq n\leq 31$. However, considering that every prime number greater than 5 can be written in the form of $6n+1$ or $6n+5$,  the following corollary shows that Theorem \ref{bound} significantly enhances the upper bound of $P(n,3)$ for all prime numbers $n\geq 37$. 
\begin{cor}\label{corollary}
Let $r\geq 6$. If $n=6r+1$ is a prime number, then 
\begin{align*}
P(n,3)&<(6r)!-(1.61)(5r+5)^{\frac{r-4}{2}},
\end{align*}
and if $n=6r+5$ is a prime number, then 
\begin{align*}
P(n,3)&<(6r+4)!-5(1.61)(5r+9)^{\frac{r-4}{2}}.
\end{align*}

\end{cor}
In Table \ref{555}, a comparison is made between the upper bounds of $P(n,3)$ obtained from \cite[Theorem 1.1]{ABJKPS} and Theorem \ref{bound} for prime numbers  $37\leq n\leq 61$.
\begin{table}[]
\begin{center}
	\begin{tabular}{|c|c|c|}
		\hline
		\cellcolor{black!20!white}{$n$}&\cellcolor{black!20!white}{\cite[Theorem 1.1]{ABJKPS}} &\cellcolor{black!20!white}{ Theorem \ref{bound}}\\
		\hline
		37 &$36!-15$&$36!-62$\\
		\hline
		41 & $40!-16$ & $40!-330$  \\
		\hline
		43 &$42!-17$&$42!-456$\\
		\hline
		47 &$46!-18$&$46!-2537$\\
		\hline
		53 &$52!-20$&$52!-155518$\\
		\hline
		59 &$58!-22$&$58!-195360$\\
		\hline
		61 &$60!-23$&$60!-323371$\\
		\hline
	\end{tabular}
	\end{center}
	\caption{\small{Comparing the upper bounds of $P(n,3)$ obtained from Theorems \cite[Theorem 1.1]{ABJKPS} and Theorem \ref{bound}. }}\label{555}
\end{table}

The subsequent sections are organized as follows: In Section \ref{prel}, we provide the definitions and notations of PCs and summarize important results regarding bounds on $P(n,d)$. Section \ref{newtable} presents a new table of values for lower bounds of $P(n,d)$ for  $n \in \{5,6,7,8\}$. In Section \ref{exact}, we first prove Theorem \ref{size4}, and subsequently, using a specific method, we determine the exact value of $ P(7,12) $. Finally, in Section \ref{secbound}, we proceed to prove Theorem \ref{bound}.

\section{Preliminaries}\label{prel}
In this section, we first present some  definitions and notations for PCs under Kendall $\tau$-metric. Subsequently, we summarize key known results about the bounds used to determine the best known bounds on PCs under Kendall $\tau$-metric in Table \ref{t1}.\\
Let $n$ be a positive integer and let $S_n$ denote the symmetric group on $n$ letters, i.e.,   the set of all $n!$ permutations of  $[n]$.  Throughout this paper, for a permutation $\pi\in S_n$, we employ the vector notation of  $\pi$ as $[\pi(1),\pi(2),\ldots ,\pi(i),\pi(i+1),\ldots ,\pi(n)]$.   The composition of two permutations   $\pi$ and  $\sigma$ in $S_n$,  denoted by $\sigma\pi$, is defined as $\sigma\pi(i) = \pi(\sigma(i))$ for all $i \in [n]$. The identity element of $S_n$ is denoted by $\xi:=[1,2,\ldots ,n]$. For distinct elements $i,j\in [n]$, $(i,j)$, which is called transposition, is  the permutation obtained  from exchanging $i$ and $j$ in $\xi$. For a permutation $\pi\in S_n$, let $I(\pi):=|\{(i,j)\in [n]^2\,|\,i<j \wedge \pi^{-1}(i)>\pi^{-1}(j)\}|$. In view of  the parity of $I(\pi)$, $\pi$ is called an even or odd permutation. For a set $Q$, $|Q|$ denotes the size of the set $Q$.\\
 For two   permutations $\pi$ and $\rho$ in $S_n$, $d_K(\rho, \pi)$ denotes  the Kendall $\tau$-distance between $\rho$ and $\pi$. There exists a  well-known equivalent expression for $d_K(\rho, \pi)$ \cite{jiang} as follows:
$
d_K(\rho, \pi)=|\{(i,j)\in[n]^2\,|\,\rho^{-1}(i)<\rho^{-1}(j)\wedge \pi^{-1}(i)>\pi^{-1}(j)\}|.
$
A PC $\mathcal{C}$ of length $n$ is called an $(n,d)$-PC, if $d_K(\pi,\sigma)\geq d$ for all distinct elements $\pi,\sigma\in \mathcal{C}$. The largest size of a $(n,d)$-PC is denoted by $P(n,d)$.
It is known that $P(n, 1)=n!$, $P(n,2)=\frac{n!}{2}$ and if $\frac{2}{3}\binom{n}{2}<d\leq \binom{n}{2} $, then $P(n,d)=2$ (see \cite[Theorem 10]{BE}). In the following, we review some results that determine the best known bounds on $P(n,d)$.\\
For a positive integer $r$ and a permutation $\sigma\in S_n$, the ball   of radius $r$ which centered at $\sigma$ in $S_n$  under the Kendall $\tau$-distance is denoted by $B_r(\sigma)$ defined by $B_r(\sigma):=\{\pi\in S_n \;|\; d_K(\sigma,\pi)\leq r\}$.   Since the Kendall $\tau$-metric is right invariant (i.e., for every three permutations $\sigma, \pi, \rho \in S_n$ we have $d_K (\sigma, \pi) = d_K (\sigma\rho,\pi\rho)$ \cite{BE}), the size of a ball of radius $r$ is independent of its center and we denote it by $B_K(r)$. The Gilbert-Varshamov bound and sphere-packing bound for PCS under Kendall $\tau$-metric are as follows:
\begin{prop}\cite[Theorems 12 and 13]{jiang}
\[
\dfrac{n!}{B_K(d-1)}\leq P(n,d) \leq \dfrac{n!}{B_K(\lfloor \frac{d-1}{2}\rfloor)}
.\]
\end{prop}
Let $\sigma$ and $\tau$ be two permutations  with $d_K(\sigma, \tau) = 1$. Then the double ball of radius $r$ centered at $\sigma$ and $\tau$, denoted by $DB_r(\sigma,\tau)$, is defined by $DB_r(\sigma,\tau):=B_r(\sigma) \cup B_r(\tau)$. The size of $DB_r(\xi,[2,1,3,\ldots ,n])$ is denoted by $DB_{n,r}$.
There are two useful results for  bounds on $P(n,d)$, when $d$ is even, as follows:
\begin{prop}\label{jb} For all $n$ and $t \geq 1$,
\begin{itemize}
\item[(1)] \cite[Corollaries 5 $\&$ 6]{BE} $P(n,2(t+1))\leq \dfrac{n!}{DB_{n,t}}$. Especially $P(n,4)\leq \dfrac{n!}{2(n-1)}$. 
\item[(2)] \cite[Theorem 21]{jiang} $P(n, 2t) \geq \frac{1}{2}P(n,2t-1)$.
\end{itemize}
\end{prop}
The best known relation for the lower bound on $P(n,3)$ is as follows:
\begin{prop}\label{case3}
$P(n,3)\geq \dfrac{n!}{2n-1} $ \cite[p. 2116]{jiang} and in particular if $n-2$ is a prime power, then $P(n,3)\geq \dfrac{n!}{2n-2} $ \cite[Theorem 4.5]{barg}.
\end{prop}
\begin{rem}
By the part (ii) of Proposition \ref{jb} and Proposition \ref{case3},  $P(n,4)\geq \dfrac{n!}{2(2n-2)}$ if $n-2$ is a prime power and $P(n,4)\geq \dfrac{n!}{2(2n-1)}$ otherwise.
\end{rem}
There is an important improvement of the lower bound on $P(n,d)$, when $n-2$ is a prime power and $d>4$ as follows:
\begin{prop}\cite[Theorem 18]{WZYG}
Let $m = ((n- 2)^{t+1} - 1)/(n - 3)$, where $n -2$ is a prime power. Then $P(n,2t+1)\geq \dfrac{n!}{(2t+1)m}$ and so $P(n,2t+2)\geq \dfrac{n!}{2(2t+1)m}$.
\end{prop}
If $\frac{1}{2}\binom{n}{2}<d\leq \frac{2}{3}\binom{n}{2}$, then the following  bound may turn out to be better than the sphere packing upper bound or part (1) of Proposition \ref{jb}.
\begin{prop}\cite[Theorem 23]{WZYG}\label{zhang}
If $P(n,2t)\geq M$, then $2\binom{M}{2}t\leq \binom{n}{2}\lfloor\frac{M}{2}\rfloor \lceil \frac{M}{2}\rceil$ and if $P(n,2t+1)\geq M$, then $(2t+2)( \binom{\lfloor\frac{M}{2}\rfloor}{2}+\binom{\lceil \frac{M}{2}\rceil}{2})+(2t+1)\lfloor\frac{M}{2}\rfloor \lceil \frac{M}{2}\rceil \leq \binom{n}{2}\lfloor\frac{M}{2}\rfloor \lceil \frac{M}{2}\rceil$. 
\end{prop}

\section{Constructing permutation codes from cosets of  subgroups}\label{newtable}
In this section, we initially devise  an algorithm that   determines the largest $(n,d)$-PC under Kendall $\tau$-metric constructed by a subgroup and some of its left cosets (see Remark \ref{cosets}, below) among all subgroups of $S_n$ for integers $n$ and $d$.
 Employing GAP \cite{gap} through this algorithm allows us to discover new $(n,d)$-PCs under Kendall $\tau$-metric, as detailed in Appendix \ref{appendix}, which improve the lower bounds of $P(n,d)$ when $n\in\{7,8\}$.  Subsequently, Table \ref{t1} is presented, illustrating the best-known bounds on $P(n,d)$ for $n\in\{5,6,7,8\}$. Recently, several improved lower bounds for $P(n, d)$ have been obtained in \cite{bereg}, using  recursive techniques,  automorphisms, and programs that combine randomness and greedy strategies.
 Notably, the bold and italic entries in the table represent  results from the current paper and \cite{bereg}, respectively. Also the blue entries shows the best known of lower bounds for $P(n, d)$, $n\in\{7,8\}$.
\begin{rem}\label{cosets}
If $H$ is a subgroup of a finite group $G$ and $g\in G$, then $Hg:=\{hg\,|\,h\in H\}$ and  $gH:=\{gh\,|\,h\in H\}$ are called a right coset of $H$ and a left
coset of $H$, respectively, with the representative $g$. It is known that if  $\textbf{X}$ be the set of right \rm{(}left\rm{)} cosets of $H$ in $G$, i.e., $\mathbf{X} := \{H g \,|\, g \in G\}$ ($\mathbf{X} := \{gH\, |\, g \in G\}$), then $\mathbf{X}$  partitions $G$, i.e., $G = \cup_{X\in \mathbf{X}} X$ and $X \cap X' = \varnothing$ for all distinct elements $X$ and $X'$ of $\mathbf{X}$, and $|\mathbf{X}|=|G|/|H|$.
\end{rem} 
\noindent\textbf{Description of Algorithm 1:} Algorithm 1 takes two input integers, $n$ and $d$. It initializes $G$ and $T$ as the symmetric group on the set $[n]$ and all subgroups of $G$, respectively (using GAP \cite{gap}, access to all subgroups of $G$ is possible with ``ConjugacyClassesSubgroups(G)"). The algorithm comprises three functions: $ \Delta $, $ \Lambda $, and $  \Theta$. The first two return the minimum Kendall $\tau$-distance between elements of a subgroup and a subset, respectively. The third function returns the minimum Kendall $\tau$-distance between $g$ and all elements of a set $M$. Notice that if $H$ is a subgroup, then since the Kendall $\tau$-metric is right invariant  and $hh_0^{-1}\in H$ for all  elements  $h,h_0\in H$, $\min\{d_K(h,h_0)\,|\,h\neq h_0,\, h,h_0\in H\}=\min\{d_K(h,\xi)\,|\, h\in H\}$. Hence, in order to reduce computer calculations, the algorithm define a separate function for calculating the  minimum kendall $\tau$-distance between elements of a subgroup.
 It initializes two lists $D$ and $L$ to be empty lists. All subgroups of $G$ that are $(n,d)$-PCs are added to the list $D$. For each $H \in D$, the algorithm initializes a list $L_H$ as the set of left transversal set $H$ in $G$ (i.e., $\{xH\, | \, x\in L(H)\}$ is the set of all left cosets of $H$ in $G$). The goal is to find the largest subset $S_H$ of $L_H$ such that $\xi\in S_H$ and $\cup_{x\in S_H}xH$ is an $(n,d)$-PC. For the latter, it first initializes two lists  $M_H$ and $S_H$ to be  list of elements of $H$ and  empty list, respectively. Next, for all $j\in L_H$, if $jH$ is an $(n,d)$-PC and  if the minimum Kendall $\tau$-distance between $j$ and all elements of $M$ is at least $d$, then it add $j$ to the set $S_H$ and $M:=M\cup jH$. Note that if there exist $x,y\in T$ such that $xH$ and $yH$ are two $(n,d)$-PC and $d_K(xh,y)\geq d $ for all $h\in H$, then the right invariant property of the Kendall $\tau$-metric implies that  $ xH\cup yH$ is an $(n,d)$-PC. With this procedure, the algorithm creates the subset $S_H$ of $L_H$ to achieve its goal. Finally, for each subgroup $H$ of $G$, it adds the list $[H,S_H]$ to $L$. Now by considering the elements of the list $L$ we can find the largest $(n,d)$-PC is created by the algorithm 1.

It is worth noting that the construction of PCs using certain subgroups of symmetric groups and their right cosets under the Hamming metric has already been explored (see, for instance, \cite{bereg1}). Since the Hamming metric on $S_n$  is left and right invariant, for adding each right (left) coset to the previously created PC of a certain subgroup and its right (left) cosets, it is enough to check the minimum distance of the representative of that coset with the previous PC.  Also, in Algorithm 1, because the Kendall $\tau$-metric is only right invariant, adding the left cosets to the previously constructed PC is used to reduce the calculations.

 \begin{table}

\begin{center}
\begin{tabular}{  m{22em} }
\hline
\textbf{Algorithm 1:} Construction $(n,d)$-PCs from the  subgroups and  some their cosets.\\
\hline\hline
\textbf{Input:} Integer numbers $n$ and  $d$.\\
\textbf{Output:} A list of elements as $[H,S_H]$ such that  $\cup_{x\in \{\xi\}\cup S_H}xH$ is an $(n,d)$-PC.\\
\\
1: $\; G \leftarrow $ symmetric group on $n$ letters\\
2: $\; T \leftarrow$  all subgroups of $G$\\
3: $\; \Delta \leftarrow$ a function whose input  is a \\ $\;\;\;\;\,$subgroup $H$\\
4: $\;\;$ $S \leftarrow []$ \\
5: $\;\;$ \textbf{for all} $i$ in $H$ \textbf{do}\\
6: $\;\;\;\;$ add Kendall $\tau$-distance between $\xi$ and $i$ to $S$\\
7: $\;\;$ \textbf{end for}\\
8: $\;\;$ return(minimum of the list $S$)\\
9: $\;$ end function\\
10: $\; \Lambda \leftarrow$ a function whose input is the \\ $\;\;\;\;\,$ subset $N$\\
11: $\;\;$ $ S \leftarrow []$ \\
12: $\;\;$ \textbf{for all} $i$ and $j$ in $N$ \textbf{do}\\
13: $\;\;\;\;\;$ add Kendall $\tau$-distance between $i$ and $j$ to $S$\\
14: $\;$ \textbf{end for}\\
15: $\;$ return(minimum of the list $S$)\\
16: $\,$ end function\\
17: $\; \Theta \leftarrow$ a function whose inputs are  a subset $M$ \\ $\;\;\;\;\,$and an element $g$ of $G$\\
18: $\;\;$ $S \leftarrow []$ \\
19: $\;\;$ \textbf{for all} $i$ in $M$ \textbf{do}\\
20: $\;\;\;\;$ add Kendall $\tau$-distance between $g$ and $i$ to $S$\\
21: $\;\;$ \textbf{end for}\\
22: $\;\;$ return(minimum of the list $S$)\\
23: $\;$ end function\\
24: $\;$ $D \leftarrow [ ]$ \\
25: $\;\;$ \textbf{for all} $i$ in $T$ \textbf{do}\\
26: $\;\;\;\;\;$ \textbf{if} $  \Delta(i)\geq d$ \textbf{then} add $i$ to the list $D$\\
27: $\;\;\;\;\;$ \textbf{end if}\\
28: $\;$ \textbf{end for}\\
29: $\;$ $L \leftarrow [ ]$ \\
30: $\;$ \textbf{for all} $H$ in $D$ \textbf{do}\\
31: $\;\;\;$ $ L_H \leftarrow $ left transversal set $H$ in $G$\\
32: $\;\;\;$ $ M_H \leftarrow $ elements of $H$\\
33: $\;\;\;$ $ S_H \leftarrow [ ]$\\
34: $\;\;\;$ \textbf{for all} $j$ in $L_H$ \textbf{do}\\
35: $\;\;\;\;\;$ \textbf{if} $ \Lambda $(elements of $(jH))\geq d$  and \\ $\;\;\;\;\,\qquad$ $\Theta(M,j)\geq d$ \textbf{then}\\ 
38: $\;\;\;\;\;\;\;$ add $j$ to the set $S_H$\\
39: $\;\;\;\;\;\;\;$ $M\leftarrow$ union of $M_H$  and the left coset $jH$ \\ $\;\;\;\;\;\;\,\qquad$of $H$ in $G$\\
40: $\;\;\;\;\;$ \textbf{end if}\\
41: $\;\;\;$ \textbf{end for}\\
42: $\;\;\;$ add $[H,S_H]$ to the set $L$\\
43: $\;$ \textbf{end for}\\
\end{tabular}
\end{center}
\end{table} 
\begin{table*}[h]
  \centering
\caption{Best known lower bound (LB) and upper bound (UP) on $P(n,d)$. }\label{t1}
\begin{tabular}{ | m{.35cm} | m{1.05cm}| m{.65cm} |m{.65cm}  |m{.65cm} |m{.65cm} |m{.65cm}| m{.65cm} |m{.65cm}  |m{.65cm} |m{.65cm} |m{.65cm}| m{.65cm} |m{.65cm}  |m{.65cm} |m{.6cm} |m{.8cm}  |} 

\multicolumn{17}{c}{}\\
  \hline
  \cellcolor{black!20!white}{$n/d$} & \cellcolor{black!20!white}{} & \cellcolor{black!20!white}{$3$} & \cellcolor{black!20!white}{ $4$}&\cellcolor{black!20!white}{$5$} &\cellcolor{black!20!white}{$6$} &\cellcolor{black!20!white}{$7$} &\cellcolor{black!20!white}{$8$} & \cellcolor{black!20!white}{$9$} &\cellcolor{black!20!white}{$10$} & \cellcolor{black!20!white}{$11$} &\cellcolor{black!20!white}{$12$} &\cellcolor{black!20!white}{$13$} &\cellcolor{black!20!white}{$14$} &\cellcolor{black!20!white}{$15$} &\cellcolor{black!20!white}{$16$} &\cellcolor{black!20!white}{17-18} \\ 
  \hline
  5 & LB=UB & $20^i$&$12^i$&$6^i$& $5^i$&$2^c$& $2^c$&$2^c$&$2^c$& --& --& --&--&--&--&--  \\
  \hline
  6 & UB \qquad LB &$116^b$  $102^i$&$64^i$  $64^i$&$26^i$  $26^i$& $20^i$  $20^i$& $11^i$  $11^i$&$7^i$ \qquad $7^i$&$4^i$ \qquad $4^i$&$4^i$ \qquad $4^i$& $2^c$ \qquad $2^c$&$2^c$ \qquad $2^c$&$2^c$ \qquad $2^c$&--&--&--&--   \\
\hline
7 & UB \qquad  \textit{LB} \qquad \textbf{LB}  \qquad  OLB & $716^b$ -- $ \qquad $  -- \qquad  $588^c$ & $420^c$ \textit{\textcolor{blue}{336}} \qquad \textbf{315}  $294^f$& $186^a$ \textit{\textcolor{blue}{126}} \qquad \textbf{\textcolor{blue}{126}} $110^d$& $120^c$    \textit{\textcolor{blue}{84}}   \qquad  \textbf{\textcolor{blue}{84}}  $55^c$& $66^a$ $ \, $ \textit{\textcolor{blue}{42}} $ \, $ \qquad\textbf{\textcolor{blue}{42}}  $34^d$& $45^c$ $ \, $ --\qquad $ \, $ \textbf{\textcolor{blue}{28}} $17^f$& $28^a$  $ \, $ -- $ \, $\qquad \textbf{\textcolor{blue}{15}} $14^d$ & $21^c$ $ \, $ \textit{\textcolor{blue}{13}} $ \, $ \qquad\textbf{12} $7^a$&$\textbf{10}$   $ \qquad $ \textit{\textcolor{blue}{8}} $ \qquad $ \textbf{\textcolor{blue}{8}} $ \qquad $   $2^a$ & $\textbf{7}$   $ \qquad $ \textit{\textcolor{blue}{7}} $ \qquad $   \textbf{\textcolor{blue}{7}}  $ \qquad $ $2^a$& $4^g$  $ \qquad $ \textit{\textcolor{blue}{4}} $ \quad $ \textbf{\textcolor{blue}{4}}$ \qquad $ $2^a$& $4^g$  $ \qquad $ -- $ \quad $ \textbf{\textcolor{blue}{4}}$ \qquad $ $2^a$ &$2^c$ $ \qquad $ -- $ \quad $ -- $ \qquad $ $2^c$&$2^c$ $ \qquad $ -- $ \quad $ -- $ \qquad $ $2^c$&$2^c$ $ \qquad $ -- $ \qquad $ -- $ \qquad $ $2^c$\\
\hline
8 & UB \qquad  \textit{LB} \qquad \textbf{LB} \qquad   OLB & $5039^c$ \textit{\textcolor{blue}{3752}} \textbf{3696}   $2688^h$ & $2880^c$ \textit{\textcolor{blue}{2240}} \textbf{2184}  $1344^a$& $1152^a$ \textit{\textcolor{blue}{672}}  \qquad\textbf{\textcolor{blue}{672}} $142^a$& $720^c$    \textit{\textcolor{blue}{448}}     \qquad\textbf{392}  $76^a$& $363^a$ $ \, $ \textit{\textcolor{blue}{168}} $ \, $ \textbf{\textcolor{blue}{168}}  $33^a$& $242^c$ $ \, $ \textit{\textcolor{blue}{115}} $ \, $ \textbf{112} $20^a$& $141^a$ $ \, $ \textit{\textcolor{blue}{57}} $ \, $ \qquad\textbf{48} $12^a$ & $99^c$ $ \, $ \textit{43} $ \, $ \qquad\textbf{\textcolor{blue}{48}} $7^a$&$64^a$ $ \qquad $ \textit{\textcolor{blue}{26}} $ \qquad $   \textbf{24}  $\qquad $   $6^a$ & $47^c$  $ \qquad $ \textit{21} $ \qquad $   \textbf{\textcolor{blue}{24}}  $ \qquad $ $4^a$& $32^a$ $ \qquad $ \textit{\textcolor{blue}{15}} $ \quad $ \textbf{14}$ \qquad $ $3^a$& $25^c$ $ \qquad $ \textit{12} $ \quad $ \textbf{\textcolor{blue}{14}}$ \qquad $ $3^a$ &$10^g$ $ \qquad $\textit{\textcolor{blue}{8}} $ \quad $ \textbf{\textcolor{blue}{8}} $ \qquad $ $1^a$&$8^g$ $ \qquad $ -- $ \quad $ \textbf{\textcolor{blue}{8}} $ \qquad $ $1^a$&$4^g$ $ \qquad $ -- $ \qquad $ \textbf{\textcolor{blue}{4}} $ \qquad $ $1^a$\\ 
\hline 
\multicolumn{17}{c}{}\\
\multicolumn{6}{c}{Key to the superscripts used in Table}&\multicolumn{11}{c}{}\\ \cline{7-17}
\multicolumn{6}{c}{}&\multicolumn{3}{c}{superscript a}&\multicolumn{8}{c}{Sphere packing bound}\\
\multicolumn{6}{c}{}&\multicolumn{3}{c}{superscript b}&\multicolumn{8}{c}{Sphere packing bound+\cite[Theorem 3.5]{ABJKPS}}\\
\multicolumn{6}{c}{}&\multicolumn{3}{c}{superscript c}&\multicolumn{8}{c}{\cite[Corollary 5 or Theorems 10,12 or 13]{BE}}\\
\multicolumn{6}{c}{}&\multicolumn{3}{c}{superscript d}&\multicolumn{8}{c}{Lower bounds from \cite{jiang}}\\
\multicolumn{6}{c}{}&\multicolumn{3}{c}{superscript f}&\multicolumn{8}{c}{\cite[Theorem 21]{jiang}}\\
\multicolumn{6}{c}{}&\multicolumn{3}{c}{superscript g}&\multicolumn{8}{c}{\cite[Theorem 23]{WZYG}}\\
\multicolumn{6}{c}{}&\multicolumn{3}{c}{superscript h}&\multicolumn{8}{c}{ $P(n,3)\geq \frac{n!}{2n-1}$ \cite{jiang}}\\
\multicolumn{6}{c}{}&\multicolumn{3}{c}{superscript i}&\multicolumn{8}{c}{\cite[Table II]{V}}\\
\multicolumn{6}{c}{}&\multicolumn{3}{c}{an entry in bold}&\multicolumn{8}{c}{Tables \ref{detail7} and \ref{detail8} and Theorem \ref{712}}\\
\multicolumn{6}{c}{}&\multicolumn{3}{c}{an entry in italic}&\multicolumn{8}{c}{Lower bounds from \cite{bereg}}\\
\multicolumn{6}{c}{}&\multicolumn{3}{c}{blue entries }&\multicolumn{8}{c}{Best known lower bounds for $P(n,d)$, $n\in \{7,8\}$}\\
\multicolumn{17}{c}{}\\
\multicolumn{17}{c}{}\\

 \end{tabular}
  
 \end{table*}

\section{The value of $P(n,d)$ for certain values of $d$}
\label{exact}

Within this section, we initially establish the proof for Theorem \ref{size4} and subsequently ascertain the exact value of $P(7,12)$. The proof of Theorem \ref{size4} relies on the following straightforward lemma.
\begin{lem}\label{partition}
Let $n\geq 5$ be an integer.  If $n\equiv 0,2$ (mod 3) {\rm(}$n\equiv 1$ (mod 3){\rm)}, then there exist 3 non-empty subsets  with the same sumset which partitions $[n]$ {\rm(}$[n]\setminus \{1\}${\rm)}, respectively.
\end{lem}
\begin{proof}
If $n$ is  5, 6, 7, 8, 9 and 10, respectively, then $\big\{\{5\},\{1,4\},\{3,2\}\big\}$, $\big\{\{6,1\},\{5,2\},$ $\{3,4\}\big\}$, $\big\{\{2,7\},\{3,6\},$ $\{4,5\}\big\}$, $\big\{\{8,4\},\{7,3,2\},\{1,5,6\}\big\}$, $\big\{\{6,5,4\},\{9,1,2,3\},$ $\{8,7\}\big\}$ and $\big\{\{10,8\},\{9,2,7\},\{3,4,6,5\}\big\}$ are the partitions of $[n]$ or $[n]\setminus\{1\}$  satisfying the lemma. Now suppose that $n>10$. Hence there exist $t>0$ and $r\in \{5,6,7,8,9,10\}$ such that $n=6t+r$. Note that if $n\equiv 1$ (mod 3), then $r\in\{7,10\}$.   Consider  $t+1$ subsets $\Theta_1$,...,$\Theta_{t+1}$ of $[n]$ as follows: 
\[
\underbrace{1,\ldots,r}_{\Theta_1},\underbrace{r+1,\ldots,r+6}_{\Theta_2},\ldots,\underbrace{n-11,\ldots,n-6}_{\Theta_{t}},\underbrace{n-5,\ldots,n}_{\Theta_{t+1}}
\]
Clearly, for all $2\leq i\leq t+1$,   $\Theta_i= \{k_i+1,k_i+2,\ldots,k_i+6\}$, where $k_i=r+(i-2)6+1$. Hence, 3 sets $\Theta_{i1}:=\{k_i+1,k_i+6\}$, $\Theta_{i2}:=\{k_i+2,k_i+5\}$ and $\Theta_{i3}:=\{k_i+3,k_i+4\}$ with the same sumset partition the set  $\Theta_i$. Therefore, since $r\in \{5,6,7,8,9,10\}$, corresponding to each set $\Theta_i$, $1\leq i\leq t+1$, there exists a partition of 3 sets $\Theta_{i1}$, $ \Theta_{i2} $ and $\Theta_{i3}$ with the same sumset for  $\Theta_i$. Let $\Delta_j:=\cup_{i=1}^{t+1} \Theta_{ij}$ for all $j\in\{1,2,3\}$. So $\Delta_1$, $\Delta_2$ and $\Delta_3$  with the same sumset partition  $[n]$ or $[n]\setminus\{1\}$ if $n\equiv 0,2$ (mod 3) or $n\equiv 1$ (mod 3), respectively. This completes the proof.
\end{proof}
\noindent\textbf{Proof of Theorem \ref{size4}.}
It  follows from \cite[Theorem 23]{WZYG} that if $P(n,d)\geq 5$, then we must have
$\binom{5}{2}d\leq 6\times \binom{n}{2}$ and therefore $d\leq \frac{3}{5}\binom{n}{2}$.  So for all $\frac{3}{5}\binom{n}{2}< d \leq \frac{2}{3} \binom{n}{2}$,  $P(n,d)\leq 4$. Since $P(n,d+1)\leq P(n,d)$, it is enough to show that there exists an $(n,\lfloor 2/3 \binom{n}{2}\rfloor)$-PC of size $4$.\\
Let   $N:=\sum_{i=1}^{n-1} i=\binom{n}{2}$. It follows from Lemma \ref{partition} that there exist  pairwise distinct subsets $\Delta_1,\Delta_2$ and $\Delta_3$ of $[n-1]$ or $[n-1]\setminus\{1\}$ such that if $n-1\equiv 0,2$ (mod 3) or $n-1\equiv 1$ (mod 3), respectively, then $\sum_{j\in \Delta_i}j=\frac{N}{3}$ or $\sum_{j\in \Delta_i}j=\frac{N-1}{3}$, for all $i\in \{1,2,3\}$. Now, suppose that for $n\geq 6$,  the  subsets $\Delta_1,\Delta_2$ and $\Delta_3$ of  $[n-1]$ are determined. Corresponding to each $\Delta_i$, we introduce a permutation $\alpha_i$ as follows: let $r_i:=|\Delta_i|$, $\Delta'_i:=\{n-j\,|\,j\in \Delta_i\}$ and $\Theta_i:=[n]\setminus \Delta'_i$.
Suppose that $j_1<j_2<\cdots <j_{r_i}$ and $l_0<l_1<\cdots <l_{n-r_i-1}$ are all elements of $\Delta'_i$ and $\Theta_i$, respectively. Let $\alpha_i\in S_n$ such that $\alpha_i(t)=j_t $ and $\alpha_i(n-s)=l_s $ for all $t\in\{1,\ldots ,r_i\}$ and $s\in\{0,\ldots ,n-r_i-1\}$. Let $\alpha_x$ and $\alpha_y$ be two distinct permutations corresponding to distinct subsets $\Delta_x$ and $\Delta_y$,  $x,y\in\{1,2,3\}$. In view of the definition of $\alpha_x$, if $i<j$ are two  elements of $[n]$, then $\alpha^{-1}_x(i)<\alpha^{-1}_x(j)$ if and only if $i\in \Delta'_x$. So, since $\Delta'_x\cap \Delta'_y=\varnothing$, we have $(i,j)\in [n]^2$ satisfies $ \alpha^{-1}_x(i)<\alpha^{-1}_x(j) $  and $ \alpha^{-1}_y(i)>\alpha^{-1}_y(j) $, if and only if  $(i,j)\in A\cup B$, where  $A:=\{(i,j)\,|\,i<j,i \in \Delta'_x\}$ and $B:=\{(i,j)\,|\,i>j,j \in \Delta'_y\}$. Hence
\begin{align*}
d_K(\alpha_x,\alpha_y)&=|\{(i,j)\,|\,  \alpha^{-1}_x(i)<\alpha^{-1}_x(j) \wedge \alpha^{-1}_y(j)>\alpha^{-1}_y(i)\}|\\&=|A\cup B|=|A|+|B|.
\end{align*}
  Therefore, $d_K(\alpha_x,\alpha_y)=\sum_{i\in \Delta_x} i+\sum_{i\in \Delta_y} i$ and so $d_K(\alpha_x,\alpha_y)$ is equal to $\frac{2N}{3} $ if $n-1\equiv 0,2$ (mod 3)  and otherwise is equal to $\frac{2(N-1)}{3}=\lfloor\frac{2}{3}N\rfloor $. Also it is easy to see that
 \begin{align*}
 d_K(\xi,\alpha_x)&=|\{(i,j)\,|\,  i<j \wedge  \alpha^{-1}_x(i)>\alpha^{-1}_x(j)\}|\\&=|\{(i,j)\,|\,i<j ,i \in \Theta_x\}|, 
 \end{align*}
and therefore $d_K(\xi,\alpha_x)$ is equal to $N-\frac{N}{3}=\frac{2}{3}N $  if $n-1\equiv 0,2$ (mod 3)  and  is equal to $N-\frac{N-1}{3}=\frac{2N+1}{3}>\lfloor \frac{2N}{3}\rfloor $ if $n-1\equiv 1$ (mod 3). Hence, $\{\xi,\alpha_1,\alpha_2,\alpha_3\}$ is an $(n,\lfloor \frac{2N}{3}\rfloor)$-PC of size $4$.   This completes the proof. \qed

\begin{ex}
Let $n=14$. Suppose   $\Delta_1:=\{2,7,8,13\}$, $\Delta_2:=\{3,6,9,12\}$ and $\Delta_3:=\{4,5,10,11\}$. Then $\Delta_1,\Delta_2,\Delta_3$ have the same sumset $30$ and   partitions  $\{2,3,\ldots ,13\}$. Hence  by the proof of Theorem{\rm\ref{size4}}, $\{\xi,\alpha_1,\alpha_2,\alpha_3\}$ is a $(14,60)$-PC, where \[\alpha_1=[1,6,7,12,14,13,11,10,9,8,5,4,3,2],\] \[\alpha_2=[2,5,8,11,14,13,12,10,9,7,6,4,3,1],\]  \[\alpha_3=[3,4,9,10,14,13,12,11,8,7,6,5,2,1].\]
\end{ex}
\begin{defn}
A permutation code  $\mathcal{C}$ is called equidistance (called EPC for short) under Kendall $\tau$-distance whenever all two distinct permutations in $\mathcal{C}$ have the same Kendall $\tau$-distance. The maximum size of the largest EPC of length $n$ and Kendall $\tau$-distance $d$ denoted by $EP(n,d)$. Also we denote by $P(n,d,m,d')$, the size of the largest PC with minimum Kendall $\tau$-distance $d$ in $S_n$ such that contains an EPC of size $m$ and Kendall $\tau$-distance $d'$.
\end{defn}
The problem of determining bounds on EPCs under the Hamming metric back to the 1970s, beginning with a question of Bolton in \cite{Bolton}. Various studies, including \cite{equi,hei, van, vans}, have explored this topic due to its applications in powerline communications and balanced scheduling. For a brief overview of EPCs under the Hamming metric, you can refer to \cite[Section VI.44.5]{col}. However there is  no special study  on EPCs under the Kendall $\tau$-metric, and only in \cite[p. 3160]{V}, the number of permutations that have the same distance with the identithy element has been studied. In the subsequent discussion, we leverage the notion of EPCs under the Kendall $\tau$-metric to demonstrate that $P(7,12)=7$.
\begin{prop}

\begin{enumerate}

\item For each $1\leq d\leq \binom{n}{2}$ and $\sigma\in S_n$, there exists an element $\pi\in S_n$ such that $d_K(\sigma,\pi)=d$.
\item If $d$ is an odd number, then $EP(n,d)=2$.
\item  $EP(n,d)=2$, for all $2/3 \binom{n}{2}<d\leq \binom{n}{2}$.
\item  $EP(n,2/3 \binom{n}{2})=4$.
\end{enumerate}

\end{prop}
\begin{proof}
First, we show that for each $1\leq t\leq \binom{n}{2}$, there exists a subset $A\subseteq [n-1]$ such that $\sum_{a\in A}a=t$. If $t< n$, then there is nothing to prove. So we assume $t\geq n$. Since $t\leq \binom{n}{2}$, there exists $i\in [n]$ such that $\sum_{j=1}^{i}(n-j)\leq t\leq \sum_{j=1}^{i+1}(n-j)$. Suppose that $s=t-\sum_{j=1}^{i}(n-j)$. So either $s=0$ or $s<n-i$. Hence, if $s=0$, then $A=\{n,n-1,\ldots ,n-i\}$ and if  $s<n-i$, then $A=\{n,n-1,\ldots ,n-i, s\}$. \\
Let $d\leq \binom{n}{2}$ and $A\subseteq [n-1]$ such that $\sum_{a\in A}a=\binom{n}{2}-d$. 
Also let $|A|=r$, $\Delta:=\{n-j\,|\,j\in A\}$ and $\Theta:=[n]\setminus \Delta$.
Suppose that $j_1<j_2<\cdots <j_{r}$ and $l_0<l_1<\cdots <l_{n-r-1}$ are all elements of $\Delta$ and $\Theta$, respectively. Let $\alpha\in S_n$ such that $\alpha(t)=j_t $ and $\alpha(n-s)=l_s $ for all $t\in\{1,\ldots ,r\}$ and $s\in\{0,\ldots ,n-r-1\}$. It is easy to see that $|\{(i,j)\in [n]^2\,|\, i<j \wedge \alpha^{-1}(i)>\alpha^{-1}(j)\}|=\binom{n}{2}-\sum_{a\in A}a=d$. Therefore $d=d_K(\xi,\alpha)=d_K(\sigma,\alpha\sigma)$ and this completes the proof of part (1). Since   the composition of two  odd (even) permutations is an even permutation, the right invariant property of  the Kendall $\tau$-metric implies that the Kendall $\tau$-distance between two permutations of the same parity is even. Hence
the part (2) follows from part (1). Also the part (3) follows from the part (1) and \cite[Theorem 10]{BE} and the part (4) follows from the proof of Theorem \ref{size4}. This completes the proof.


\end{proof}
\begin{lem}\label{pe}
$EP(7,12)=7$ and  $P(7,11,6,12)=7$.
\end{lem}
\begin{proof}
Let $\mathcal{C}$ be an $(7,12)-$EPC under the Kendall $\tau-$ metric of  maximum size. Without loss of generality we may assume  that $\xi\in \mathcal{C}$ as  Kendall $\tau$-metric is right invariant. Let $\mathcal{A}:=\{\sigma\in S_n\,|\, d_K(\xi,\sigma)=12\}$. Using GAP \cite{gap}, $|\mathcal{A}|=531$. It is sufficient that we find the maximum EPC with  Kendall $\tau$-distance 12 in $A$. Let $\mathcal{A}_i$, $2\leq i\leq 7$, be the set of all subsets of size $i$ in $\mathcal{A}$ such that the  Kendall $\tau$-distance between any pair of distinct elements in any of them is equal to 12. Using GAP, $|\mathcal{A}_2|=27697$, $|\mathcal{A}_3|=172629$, $|\mathcal{A}_4|=131777$, $|\mathcal{A}_5|=10862$, $|\mathcal{A}_6|=9$ and $|\mathcal{A}_7|=0$. So, the size of largest EPC in $\mathcal{A}$ is 6 and therefore $|\mathcal{C}|=7$.\\
Suppose that $\mathcal{C}$ is a $(7,11)$-PC such that contains an EPC $\bar{\mathcal{C}}$ of size $6$ and the Kendall $\tau$-distance $12$. Without loss of generality we may assume that $\xi\in \bar{\mathcal{C}}$. According to the proof of the first part, $\bar{\mathcal{C}}\in \mathcal{M}:= \{\{\xi\}\cup A\,|\,A \in \mathcal{A}_5\} $. So there are 10862 distinct cases for $\bar{\mathcal{C}}$. Let $\mathcal{B}_M:=\{\sigma\in S_n\,|\, d_K(m,\sigma)\geq 11, \forall \,\,m\in M\}$, for all $M\in \mathcal{M}$. Using GAP, for all $M\in \mathcal{M}$, $0\leq |\mathcal{B}_M|\leq 14$ and if $|\mathcal{B}_M|\neq 0$ and  $b_1,b_2\in \mathcal{B}_M$ then  $d_K(b_1,b_2)<11$. This completes the proof. 
\end{proof}
\begin{thm}\label{712}
 $P(7,12)=7$ and $8\leq P(7,11)\leq 10$.
\end{thm}
\begin{proof}
  Let $\mathcal{C}$ be a $(7,d)-$PC under the Kendall $\tau-$metric and  $\Sigma:=\sum_{c_1,c_2\in \mathcal{C}}d_K(c_1,c_2)$.   By a same  argument as in the proof of \cite[Theorem 23]{WZYG}, it can be  seen that $\Sigma\leq \binom{n}{2}\lceil\frac{|\mathcal{C}|}{2}\rceil \lfloor\frac{|\mathcal{C}|}{2}\rfloor$. By Theorem \ref{zhang}, $P(7,12)\leq 8$. As shown in Table \ref{t1}, $P(7,12)\geq 7$.   Then it is sufficient that we show  $P(7,12)\neq 8$. Suppose for a contradiction that $\mathcal{C}$ be an $(7,12)-$PC of  size 8. So we must have $\Sigma\leq 336$. On the other hand, since $EP(7,12)=7$ and $|\mathcal{C}|=8$, there exist $c_1,c_2$ in $\mathcal{C}$ such that $d_K(c_1,c_2)>12$. Hence, $\Sigma\geq \binom{8}{2}\times 12+1=337$ that is a contradiction. So $P(7,12)=7$.\\ 
As shown in Table \ref{t1},  $P(7,11)\geq 8$. Theorem \ref{zhang} implies that $P(7,11)\leq 12$. Suppose for a contradiction that $\mathcal{C}$ be an $(7,11)-$PC of  size 12 or 11. Let $\mathcal{C}_1:=\mathcal{C}\cap A_n$ and $\mathcal{C}_2:=\mathcal{C}\backslash \mathcal{C}_1$, where $A_n$ denotes the set of all even permutation in $S_n$. Without loss of generality we may assume that $|\mathcal{C}_1|\geq |\mathcal{C}_2|$. Since the Kendall $\tau$-distance between two permutations of the same parity is even, if $c_1$ and $c_2$ are two distinct elements in $\mathcal{C}_1$ or $\mathcal{C}_2$, then $d_K(c_1,c_2)\geq 12$. So
\[ \binom{|\mathcal{C}_1|}{2}\times 12+ 
\binom{|\mathcal{C}_2|}{2}\times 12+11\times |\mathcal{C}_1|\times |\mathcal{C}_2|\leq \Sigma.\]
If $|\mathcal{C}|=12$ and 11, then  $\Sigma\leq 756$ and 630, respectively. Hence it can  be  seen that if $|\mathcal{C}|=12$, then  $\mathcal{C}_1$ and $\mathcal{C}_2$ must be two $(7,12)-$EPC of sizes 6 and also if $|\mathcal{C}|=11$, then  $\mathcal{C}_1$ and $\mathcal{C}_2$ must be  two $(7,12)-$EPC of sizes 6 and 5, respectively. Therefore if $|\mathcal{C}|\in\{11,12\}$, then $\mathcal{C}$ is an $(7,11)-$PC such that contains a $(7,12)-$EPC of size 6 that is  contradict with Lemma \ref{pe}, This completes the proof.
\end{proof}
\section{New upper bound of $P(n,3)$}\label{secbound}
In this section, we will follow the definitions and notations outlined in \cite{ABJKPS}. Specifically, we adopt the following definitions:
Let $H$ be a subgroup of a finite group $G$ and $X=\{Ha_1,\ldots ,Ha_m\}$ be the set of right cosets of $H$
in $G$. Fix  the ordering on $X$ as  $H a_i < H a_j$ whenever $i < j$. Then $\rho_X^G$ is the
map from $G$ to $GL_m(\mathbb{Z})$ (the group of all $m \times m$ invertible matrices with integer entries)
defined by $g \rightarrow P_g$, where $P_g$ is the  $m \times m$ matrix whose $(i, j)$ entry is 1 if $Ha_ig=Ha_j$
 and 0 otherwise. Furthermore, if $Y\subseteq G$, then $ \widehat{Y^{\rho_{X}^{G}}} $ represents the element $\sum_{y\in Y}y^{\rho_{X}^{G}}=\sum_{y\in Y}P_y$ in $Mat_m (\mathbb{Z})$, the set of all $m \times m$ matrices over $\mathbb{Z}$. \\
By a number  partition $\lambda$ of $n$ {\rm(}with the length $m)$ we mean an $m$-tuple $(\lambda_1,\dots,\lambda_m)$ of positive integers such that $\lambda_1\geq \cdots\geq\lambda_m$ and $n=\sum_{i=1}^m \lambda_i$. According to \cite[Definition 3.1 and Remark 3.2]{ABJKPS}, the Young subgroup corresponding to a partition $(\lambda_1,\dots,\lambda_m)$ of a positive integer $n$ refers to the subgroup $H$ of $S_n$ defined as $H :=S_{\Delta_1}\times \cdots\times S_{\Delta_m}= \{\sigma_1 \cdots  \sigma_m \,|\, \sigma_i \in \Delta_i, 1 \leq i \leq m\}$, where $(\Delta_1,\dots,\Delta_m) = (\{1,\ldots ,\lambda_1\},\{\lambda_1+1,\ldots ,\lambda_1+\lambda_2\},\ldots,\{n-\lambda_m+1,\ldots ,n\})$, and $S_{\Delta_i}$ denotes the symmetric group on the set $\Delta_i$ for all $i = 1,\ldots ,m$. 
\begin{lem}\label{matrix}
Let   $H$ be the Young subgroup of $S_n$ corresponding to the partition $\lambda:=(n-r,\underbrace{1, \ldots ,1}_r)$ and $X$ be the set of  right cosets of $H$ in $S_n$. If $S=\{(i,i+1)\,|\,1\leq i\leq n-1\}$ and $T:=S\cup \{\xi\}$,  then $ \widehat{T^{\rho_{X}^{S_n}}} $ is a symmetric  matrix $A=(a_{ij})_{\ell\times \ell}$, where $\ell=\frac{n!}{(n-r)!}$, with   the following properties:
\begin{enumerate}
\item  $a_{ii}\geq n-2r$ for all $i\in [\ell]$. 
\item  $a_{ij}\in\{0,1\}$ for all $i\neq j\in [\ell]$.
\item $\sum_{j=1}^{\ell}a_{ij}=n$ for all $i\in [\ell]$. 
\end{enumerate}
\end{lem}
\begin{proof}
In view of \cite[Remark 3.2]{ABJKPS}, without loss of generality we may assume that $\lambda$ is the partition $\{[n-r],\{n-r+1\},\{n-r+2\},\ldots ,\{n\}\}$ of $n$ and therefore $H\cong S_{n-r}$. Let $\mathcal{F}:=\{(f_1,f_2,\ldots ,f_r)\in [n]^r\,|\, \forall\, i\neq j, f_i\neq f_j\}$. Corresponding to each  ordered $r$-tuple $F=(f_1,\ldots ,f_r)\in \mathcal{F}$, let $S_n^F:=\{\sigma\in S_n\,|\,\sigma(n-r+1)=f_1,\sigma(n-r+2)=f_2,\ldots ,\sigma(n)=f_r\}$.
Clearly, $S_n^F=Hg$, where $g=(n-r+1,f_1)(n-r+2,f_2)\cdots (n,f_r)$, is a right coset of $H$ and if $F$ and $\bar{F}$ are two distinct elements of $\mathcal{F}$, then  $S_n^{F}\cap S_n^{\bar{F}}=\varnothing$. Hence, since $|\mathcal{F}|=\ell$, in view of Remark \ref{cosets}, $X=\{S_n^F\,|\,F\in \mathcal{F}\}$ is the set of all right cosets of $H$ in $S_n$. Suppose that $F_1,F_2,\ldots ,F_{\ell}$ are all ordered $r$-tuples in $\mathcal{F}$.
 Fix the ordering of $X$ such that $S_n^{F_i}<S_n^{F_j}$ if $i<j$, for all $i,j\in [\ell]$. In view of \cite[Definition 2.10]{ABJKPS}, the $(i,j)$ entry of $ \widehat{T^{\rho_{X}^{S_n}}} $ is equal to $|\mathcal{O}_{ij}|$, where $\mathcal{O}_{ij}:=\{t\in T\,|\,S_n^{F_i}t=S_n^{F_j}\}$. Since $\mathcal{O}_{ij}=\mathcal{O}_{ji}$ for all $i,j\in[\ell]$, $A$ is a symmetric matrix.  Let $(i,i+1)\in T$ and let $F=(f_{1},\ldots ,f_{r})$ and $\bar{F}=(\bar{f}_{1},\ldots ,\bar{f}_{r})$ be two distinct elements of $\mathcal{F}$. The sufficient condition for $S_n^{F}(i,i+1)=S_n^{F}$ is $\{i,i+1\}\cap \{f_{1},\ldots ,f_{r}\}=\varnothing$. So $a_{ss}\geq n-2r$ for all $s\in [\ell]$. Now suppose for a contradiction that there exists $(j,j+1)\in T\setminus \{(i,i+1)\}$ such that $S_n^{F}(i,i+1)=S_n^{\bar{F}}=S_n^{F}(j,j+1)$. Since $F\neq  \bar{F}$ we have $P_1:=\{f_{1},\ldots ,f_{r}\}\cap \{i,i+1\}\neq \varnothing$ and $\{f_{1},\ldots ,f_{r}\}\cap \{j,j+1\}\neq \varnothing$. Suppose that $i\in P_1$ and $f_{m}=i$ for some $m\in [r]$.
 Then for all $\sigma\in S_n^F$,  $\big(\sigma (i,i+1)\big)(n-r+m)=i+1$ and  $ \big(\sigma (j,j+1)\big)(n-r+m)$ is equal to $j$ if $i=j+1$ and  is equal to $i$ if $\{i,i+1\}\cap \{j,j+1\}=\varnothing$. So $S_n^{F}(i,i+1)\neq S_n^{F}(j,j+1)$  that is a contradiction. Now suppose that $i+1\in P_1$ and $f_{d}=i+1$ for some $d\in [r]$.
 then by the same argument it can be seen that  $\big(\sigma (i,i+1)\big)(n-r+d)\neq \big(\sigma (j,j+1)\big)(n-r+d)$, for all $\sigma\in S_n^F$ . Hence,  $S_n^{F}(i,i+1)\neq S_n^{F}(j,j+1)$  that is a contradiction.
 Hence, $a_{ij}\in\{0,1\}$ for all $i\neq j\in [\ell]$. Note that for each $x\in[\ell]$, since $\cup_{y=1}^{\ell}\mathcal{O}_{xy}=T$ and $\mathcal{O}_{xy}\cap \mathcal{O}_{xy'}=\varnothing$ for all $y\neq y'\in [\ell]$, we have $\sum_{j=1}^{\ell}a_{ij}=n$ for all $i\in [\ell]$. 
 This completes the proof.
\end{proof}
Here, we provide some notations used in the proof of Theorem \ref{bound}. The transposition of a matrix or vector is denoted by $ \cdot^t $. The inner product of two vectors $\textbf{x}=(x_1,\ldots,x_n)^t$ and $\textbf{y}=(y_1,\ldots,y_n)^t$ in $\mathbb{R}^n$ is defined as $\left\langle  \textbf{x}, \textbf{y}\right\rangle  :=\textbf{x}^t\textbf{y}=\sum_{i=1}^{n}x_iy_i$, the notation $\parallel \textbf{x}\parallel:=\sqrt{\left\langle  \textbf{x},\textbf{x}\right\rangle }$ denotes the 2-norm of vector $\textbf{x}$ and the notation $\parallel \textbf{x}\parallel_1:=\sum_{i=1}^n|x_i|$ denotes the 1-norm of vector $\textbf{x}$, where $|a|$ denotes the absolute value of real number $a$.  
\begin{defn}\label{poly}
\cite{Sch} A polyhedral cone is a subset $\mathcal{C}\subset\mathbb{R}^n$ of the form $\mathcal{C}:=\{\textbf{x}\in \mathbb{R}^n\,|\, A\textbf{x}\leq \textbf{0}\}$,
for a matrix $A\in R^{m\times n}$ and column vector $\textbf{0}$ of order $n\times 1$ whose entries are equal to $0$. 
\end{defn}
 \begin{rem}\label{exterme}
 Let $\mathcal{C}=\{\textbf{x}\in \mathbb{R}^n\,|\, A\textbf{x}\leq \textbf{0}\}$ be a polyhedral cone for a non-singular matrix $A\in \mathbb{R}^{n\times n}$ . In view of \cite[p. 104-105]{Sch}, the vector $\textbf{d}\in \mathbb{R}^{n}$ is called an extreme ray of $C$, if there exists $1\leq i\leq n$ such that  $A_i \textbf{d}=\textbf{0}$ and $a_i\textbf{d}\leq 0$, where   $a_i$ denotes the $i$-th row of the matrix $A$ and $A_i$ is the submatrix  of $A$ obtained by removing $a_i$.   We say that two extreme rays $\textbf{d}$  and $\textbf{d}'$  of $\mathcal{C}$ are equivalent, and denote it by $\textbf{d} \sim \textbf{d}'$, if one is a positive multiple of the other.  In view of \cite[p. 101-105]{Sch},  the number of  equivalence classes of extreme rays in $\mathcal{C}$ is finite. Also according to \cite[p. 105]{Sch}, if $\{\textbf{w}_1,\ldots , \textbf{w}_s\}$ is a complete set of  representatives of all equivalence classes of extreme rays in $\mathcal{C}$, then $\mathcal{C}=\{\sum_{i=1}^{s}\lambda_i\textbf{w}_i \,|\, \lambda_i\geq 0\}$.
 \end{rem}


  
\begin{thm}\label{bound1}
Let $r$ and $n$ be integers such that $r< \frac{n}{4}$ and $n\nmid (n-r)!$. Then 
\[
P(n,3)\leq (n-1)!-\dfrac{n-6r}{\sqrt{n^2-8rn+20r^2}}\sqrt{\dfrac{(n-1)!}{n(n-r)!}}.
\]
\end{thm}
\begin{proof}
Let $C$ be a PC in $S_n$ with minimum Kendall $\tau$-distance 3. Let $H$ be the Young subgroup of $S_n$ corresponding to the partition $\lambda:=(n-r,\underbrace{1, \ldots ,1}_r)$ and $Y$ be the set of  right cosets of $H$ in $S_n$. If $S=\{(i,i+1)\,|\,1\leq i\leq n-1\}$ and $T:=S\cup \{\xi\}$,  then by Lemma \ref{matrix}, $ \widehat{T^{\rho_{Y}^{S_n}}} $ is a   matrix $A=(a_{ij})_{\ell\times \ell}$, $\ell=\frac{n!}{(n-r)!}$, with properties specified in Lemma \ref{matrix}.  Theorem \cite[2.14]{ABJKPS} implies that the optimal value of the objective function of the following integer programming problem  gives an upper bound on $|C|$
\begin{align*}
\text{Maximize} &\quad \sum_{i=1}^{\ell}{x_i},\\
\text{subject to}&\quad A(x_1,\ldots ,x_{\ell})^t \leq |H| \mathbf{1}=(n-r)!\mathbf{1},\\
& \quad x_i\in \mathbb{Z},\,\, x_i\geq 0, \,\, i\in\{1,\ldots,\ell\},
\end{align*}
where $\textbf{1}$ is a column vector of order $\ell\times 1$ whose  entries are equal to $1$. Let $\pmb{\alpha}$ be a feasible solution for the above linear inequality system  that achieves the optimum of the objective function and   $\pmb{\beta}:= \frac{(n-r)!}{n}\mathbf{1}$. It follows from the part (3) of Lemma \ref{matrix} that  the sum of every row in $A$ is equal to $n$ and  so $A\pmb{\beta}=(n-r)!\mathbf{1}$. Since $n\nmid (n-r)!$  we have $\pmb{\alpha}\neq \pmb{\beta}$. It is clear that $\sum_{i=1}^{\ell}\alpha_i\leq (n-1)!$, where    $\alpha_{i}$ denotes the $i$-th entry of  $\pmb{\alpha}$, and suppose that  $\sum_{i=1}^{\ell}\alpha_i=(n-1)!-k$ for a non-negative integer $k$.  Consider two vectors $\overrightarrow{\pmb{\beta}\pmb{\alpha}}:=\pmb{\alpha}-\pmb{\beta}$ and  $-\textbf{1}$. We let 
\begin{align*}
\mu &:=\dfrac{\left\langle -\textbf{1},\overrightarrow{\pmb{\beta}\pmb{\alpha}}\right\rangle }{\parallel -\textbf{1}\parallel \parallel \overrightarrow{\pmb{\beta}\pmb{\alpha}}\parallel }=\dfrac{\left\langle -\textbf{1},\pmb{\alpha}-\pmb{\beta} \right\rangle }{\parallel -\textbf{1}\parallel \parallel \pmb{\alpha}-\pmb{\beta}\parallel }\\ &=
\dfrac{\left\langle -\textbf{1},\pmb{\alpha}\right\rangle + \left\langle -\textbf{1},-\pmb{\beta}\right\rangle }{\parallel -\textbf{1}\parallel \parallel \pmb{\alpha}-\pmb{\beta}\parallel }\\ &=\dfrac{\ell\frac{(n-r)!}{n}-\sum_{i=1}^{\ell}\alpha_i}{\sqrt{\ell}\sqrt{\sum_{i=1}^{\ell}(\alpha_{i}-\beta_{i})^2}}\\&=\dfrac{k}{\sqrt{\ell}\sqrt{\sum_{i=1}^{\ell}(\alpha_{i}-\beta_{i})^2}},
\end{align*}
where $\beta_{i}$  denotes the $i$-th entry of $\pmb{\beta}$. Since for each $i\in [\ell]$, $\alpha_{i}$ is an integer number,
we have  $|\alpha_{i}-\beta_{i}|\geq \frac{1}{n}$. Hence,
\begin{equation}\label{eq0}
k\geq \mu \sqrt{\ell}\sqrt{\frac{\ell}{n^2}}=\mu\frac{\ell}{n}=\dfrac{(n-1)!}{(n-r)!}\mu.
\end{equation}
Let $\mathcal{C}:=\{\mathbf{x}\in \mathbb{R}^{\ell}\,|\,A\mathbf{x} \leq (n-r)!\mathbf{1}\}=\{\mathbf{x}\in \mathbb{R}^{\ell}\,|\,A(\mathbf{x}-\pmb{\beta}) \leq \mathbf{0}\}$. In  view of Definition \ref{poly}, $\mathcal{C}$ is a polyhedral cone.  Note that since $r< \frac{n}{4}$,  Lemma \ref{matrix} implies that $A=(a_{ij})_{\ell\times \ell}$ is a matrix such that $a_{ii}>\sum_{i\neq j=1}^{\ell}a_{ij}$ for all $1\leq i\leq \ell$.  Therefore  Levy-Desplanques Theorem \cite[p. 125]{horn} implies $A$ is a non-singular matrix. Also, since $\lambda_0 \mathbf{u}+(1-\lambda_0)\mathbf{v}\in \mathcal{C}$ for all $\mathbf{u},\mathbf{v}\in \mathcal{C}$ and $\lambda_0\in [0,1]$, $\mathcal{C}$ is a convex set. It is clear that $\pmb{\beta},\pmb{\alpha}\in \mathcal{C}$ and so  the vector $\overrightarrow{\pmb{\beta}\pmb{\alpha}}$ belongs to $\mathcal{C}$.
 Suppose that $\{\textbf{w}_1,\ldots , \textbf{w}_s\}$ is a complete set of  representatives of all equivalence classes of extreme rays in $\mathcal{C}$ such that $\parallel \textbf{w}_i\parallel=1$ for all $1\leq i\leq s$.  Since $\overrightarrow{\pmb{\beta}\pmb{\alpha}}\in \mathcal{C}$, it follows from  Remark \ref{exterme} that  there exist non-negative real numbers $\lambda_1, \ldots ,\lambda_s$ such that $\overrightarrow{\pmb{\beta}\pmb{\alpha}}=\sum_{i=1}^{s}\lambda_i\textbf{w}_i$.  Then
\begin{align*}
\mu &=\dfrac{\left\langle -\textbf{1},\overrightarrow{\pmb{\beta}\pmb{\alpha}}\right\rangle }{\parallel \textbf{1}\parallel \parallel \overrightarrow{\pmb{\beta}\pmb{\alpha}}\parallel }=\dfrac{\left\langle -\textbf{1},\sum_{i=1}^{s}\lambda_i\textbf{w}_i\right\rangle }{\parallel -\textbf{1}\parallel \parallel\sum_{i=1}^{s}\lambda_i\textbf{w}_i\parallel}. 
\end{align*}
Since $\parallel \sum_{i=1}^{s}\lambda_i\textbf{w}_i \parallel \leq \sum_{i=1}^{s}\lambda_i\parallel \textbf{w}_i \parallel$, 
\begin{align*}
\mu&\geq \dfrac{\sum_{i=1}^{s}\lambda_i\left\langle -\textbf{1},\textbf{w}_i\right\rangle }{\parallel -\textbf{1}\parallel (\sum_{i=1}^{s}\lambda_i\parallel \textbf{w}_i\parallel ) }, 
\end{align*} 
and since $\parallel \textbf{w}_i\parallel=1$ for all $1\leq i\leq s$, 
\begin{align}
\mu&\geq \sum_{i=1}^{s}\dfrac{\lambda_i\left\langle -\textbf{1},\textbf{w}_i\right\rangle }{ (\sum_{j=1}^{s}\lambda_j)\parallel -\textbf{1}\parallel  \nonumber }\\  &= \sum_{i=1}^{s}\dfrac{\lambda_i}{\sum_{j=1}^{s}\lambda_j}\dfrac{\left\langle -\textbf{1},\textbf{w}_i\right\rangle }{ \parallel-\textbf{1}\parallel   } \geq \sum_{i=1}^{s}\dfrac{\lambda_i}{\sum_{j=1}^{s}\lambda_j}\mu_0=\mu_0, \label{eq11}
\end{align}
where  $\mu_0:=\min\left\{\dfrac{\left\langle -\textbf{1},\textbf{w}_i\right\rangle}{\parallel -\textbf{1}\parallel}\, \big |\,1\leq i\leq s\right\}$.\\
  Suppose  that $\mu_0=\dfrac{\left\langle -\textbf{1},\textbf{w}_r\right\rangle }{ \parallel-\textbf{1}\parallel   }$ for some $1\leq r\leq s$. Hence it follows from Remark  \ref{exterme} that there exists $i\in [n]$ such that  $A_i\textbf{w}_r=\textbf{0}$ and $a_i\textbf{w}_r\leq 0$, where $a_i$ is the $i$-th row of the matrix $A$ and  $A_i$ is the matrix obtained by removing $a_i$ of the matrix $A$. According to the properties of the matrix $A$, without loss of generality, we may assume that $i=\ell$. 
  Suppose that $\pmb{\rho}$ is the $\ell$-th column of $A_{\ell}$ and $J$ is the $(\ell-1)\times (\ell-1)$ matrix obtained by removing the column $\pmb{\rho}$  of the matrix $A_{\ell}$.   Levy-Desplanques Theorem  implies $J$ is a non-singular matrix. Hence, $A_{\ell}(x_1,\ldots ,x_{\ell})^t= J(x_1,\ldots  ,x_{\ell-1})^t+\pmb{\rho}x_{\ell}=\textbf{0}$ implies  $(x_1,\ldots  ,x_{\ell-1})^t=-J^{-1}\pmb{\rho}x_{\ell}$.\\ In the sequal, we show that $a_{\ell}(J^{-1}\pmb{\rho},-1)^t\leq 0$  and therefore by placing $x_{\ell}=-1$ we have  $(-J^{-1}\pmb{\rho}x_{\ell},x_{\ell})^t\sim\textbf{w}_r$.  It follows from \cite[Theorem 1]{varah} and Lemma \ref{matrix} that if  $\Delta:=\min\{ |J_{ii}|-\sum_{ j=1,j\neq i}^{\ell-1}|J_{ij}|\,|\,1\leq i\leq \ell-1\}$, then $\parallel J^{-1}\parallel_{\infty}:=\max\{\sum_{ j=1}^{\ell-1}|(J^{-1})_{ij}|\,|\,1\leq i\leq \ell-1\}\leq \frac{1}{\Delta}$. So Lemma \ref{matrix} implies $\parallel J^{-1}\parallel_{\infty}\leq \frac{1}{n-4r}$. Also if $|A|:=(|a_{ij}|)_{n\times n}$  for a matrix $A=(a_{ij})_{n\times n}$, then we have 
  \begin{align*}
  \parallel J^{-1}\pmb{\rho} \parallel_1 =&\tr(|J^{-1}\pmb{\rho}|\textbf{1}) \leq \tr(|J^{-1}|\pmb{\rho}\textbf{1}).
\end{align*}
  Since the inverse of a symmetric matrix is a symmetric matrix, $J^{-1}$ is a symmetric matrix. Suppose that $\rho_i$ denotes the $i$-th entry of $\pmb{\rho}$. It follows from Lemma \ref{matrix} that  $\rho_i\in\{0,1\}$ for all $1\leq i\leq \ell -1$  and if  $\tau:=\{i\in [\ell -1]\, |\, \rho_i=1\}$, then  the size of $\tau$ is at most $ 2r$.   Then we have
  \begin{align*}
  \tr(|J^{-1}|\pmb{\rho}\textbf{1})&=\sum_{i=1}^{\ell -1}\sum_{j\in \tau}|(J^{-1})_{ij}|\\ &=\sum_{j\in \tau}\sum_{i=1}^{\ell -1}|(J^{-1})_{ij}|\\ &=\sum_{j\in \tau}\sum_{i=1}^{\ell -1}|(J^{-1})_{ji}|\\ &\leq \sum_{j\in \tau}\parallel J^{-1}\parallel_{\infty}\leq 2r \parallel J^{-1}\parallel_{\infty},
  \end{align*}
 and therefore,
\begin{align}
  \parallel J^{-1}\pmb{\rho} \parallel_1   \leq \frac{2r}{n-4r}.\label{eq21}
\end{align}
  So, parts (1) and (2) of Lemma \ref{matrix} and $r<\dfrac{n}{4}$ imply that $$a_{\ell}(J^{-1}\pmb{\rho},-1)^t \leq \parallel J^{-1}\pmb{\rho} \parallel_1 -(n-2r) \leq 0$$ 
  and so $(J^{-1}\pmb{\rho},-1)^t\sim\textbf{w}_r$.     Hence,    
\begin{align}
\mu_0&=\dfrac{\left\langle  -\textbf{1}, (J^{-1}\pmb{\rho},-1)^t\right\rangle }{\parallel \textbf{1}\parallel \parallel (J^{-1}\pmb{\rho}, -1)^t \parallel}=\dfrac{1-\left\langle  \textbf{1}, J^{-1}\pmb{\rho}\right\rangle }{\sqrt{\ell}\sqrt{1+\parallel J^{-1}\pmb{\rho} \parallel^2}\nonumber} \\ \label{eq31} &\geq  \dfrac{1- \parallel J^{-1}\pmb{\rho} \parallel_1}{\sqrt{\ell}\sqrt{1+\parallel J^{-1}\pmb{\rho} \parallel_1^2}}.
\end{align}
Hence, relations \eqref{eq21} and \eqref{eq31} imply 
\begin{equation}\label{end}
\mu_0\geq \dfrac{n-6r}{\sqrt{\ell}\sqrt{n^2-8rn+20r^2}},
\end{equation}
  and therefore the result follows from relations \eqref{eq0}, \eqref{eq11} and \eqref{end}. This completes the proof.
\end{proof}
\begin{prop}\label{primecase}
For integers $n\geq 10$ and $r\leq \dfrac{n}{2}$, if $n\nmid (n-r)!$, then $n$ is a prime number.
\end{prop}
\begin{proof}
Suppose for a contradiction that $n$ is not prime. Hence there exist $n_1,n_2\in \mathbb{N}\setminus\{1\}$ such that $n=n_1n_2$. Suppose first that $n_1\neq n_2$ and  $n_1<n_2$. If $n_2\leq n-r$, then $n| (n-r)!$ that is a contradiction. So $n_2> n-r$. Since $r\leq \frac{n}{2}$, $$\frac{n}{2}\leq n-r <n_2=\frac{n}{n_1},$$ and therefore $n_1<2$ that is a contradiction. Now suppose that $n=n_1^2$. Since $n\nmid (n-r)!$, $n-r< 2n_1$ and so
\[\frac{n_1^2}{2}= \frac{n}{2}\leq n-r <2n_1,
\]
and therefore $n_1<4$ that is a contradiction. This completes the proof.
\end{proof}
\begin{rem}
In view of Proposition \ref{primecase}, the only numbers that satisfy the assumptions of Theorem \ref{bound1} are prime numbers. Thus,  given that $P(n,3)\leq (n-1)!$, Theorem \ref{bound1} is interchangeable with Theorem \ref{bound}.
\end{rem}
\noindent\textbf{Proof of Corollary \ref{corollary}:} First suppose that $n=6r+1\geq 37$ is a prime number. It follows from inequality \ref{mainrelation} that 
\begin{align*}
P(6r+1,3)&\leq (6r)!-\sqrt{\frac{(6r)(6r-1)\cdots (5r+2)}{48r^3+32r^2+10r+1}}.
\end{align*}
Clearly,  $\dfrac{(5r+2)(5r+3)(5r+4)}{48r^3+32r^2+10r+1}> 2.6$. Therefore, 
\begin{align*}
P(6r+1,3)&<(6r)!-\sqrt{2.6}(5r+5)^{\frac{r-4}{2}}\\&<(6r)!-\lceil\frac{6r+1}{3}\rceil+2.
\end{align*}
Now suppose that $n=6r+5\geq 41$ is a prime number. It follows from inequality \ref{mainrelation} that  
\begin{align*}
P(6r+5,3)&\leq (6r+4)!-5\sqrt{\frac{(6r)(6r-1)\cdots (5r+6)}{48r^3+160r^2+250r+125}}.
\end{align*}
Clearly, $\dfrac{(5r+6)(5r+7)(5r+8)}{48r^3+160r^2+250r+125}> 2.6$. Therefore, 
\begin{align*}
P(6r+5,3)&<(6r+4)!-5\sqrt{2.6}(5r+9)^{\frac{r-4}{2}}\\&<(6r+4)!-\lceil\frac{6r+5}{3}\rceil+2.
\end{align*}
This completes the proof.\qed



%





\ifCLASSOPTIONcaptionsoff
  \newpage
\fi



%







\section{Appendix}\label{appendix}
Tables \ref{detail7} and \ref{detail8} contain information on the generators of subgroup $H$ in $S_n$ (i. e., a subset of the elements $H$  such that every element of the $H$ can be expressed as a combination  of finitely many elements of the subset and their inverses) and $S_H$ for $n\in\{7,8\}$ as obtained from Algorithm 1, along with additional software checking details. In these tables, $C_n^d$ indicates the size of the set of all subgroups of $S_n$ that are $(n,d)$-codes under Kendall $\tau$-metric and $\lambda_H$ indicates the number of left cosets of $H$ that are $(n,d)$-codes under Kendall $\tau$-metric.  In fact in these tables, for every pair $(n,d)$,  $\cup_{x\in S_H\cup {\xi}}xH$ forms a new $(n,d)$-PC.

\begin{table}[h]

\begin{center}
\begin{tabular}{ | m{.35cm} | m{.75cm}| m{6.7em} |m{.35cm}  |m{.45cm} |m{6.7em} |}

  \hline
  \cellcolor{black!20!white}{$d$} & \cellcolor{black!20!white}{$|C_7^d|$} & \cellcolor{black!20!white}{Generators of $H$}&  \cellcolor{black!20!white}{$|H|$}&\cellcolor{black!20!white}{$\lambda_H$} & \cellcolor{black!20!white}{Elements of $S_H$}\\ 
  \hline
 \cellcolor{black!20!white}{ $4$} & $5565$ & $ [ 4, 2, 1, 3, 7, 5, 6 ]$  $ [ 3, 5, 2, 6, 7, 1, 4 ] $&  $21$&$240$ & $[ 1, 2, 3, 7, 4, 6, 5 ]$ $ [ 1, 2, 3, 5, 7, 6, 4 ]$  $ [ 1, 2, 6, 3, 4, 7, 5 ]$ $  [ 1, 2, 4, 7, 3, 5, 6 ]$  $
  [ 1, 2, 5, 4, 3, 7, 6 ]$ $  [ 1, 2, 5, 6, 3, 4 ,7]$  $ [ 1, 2, 4, 6, 5, 3,7 ] $ $ [ 1, 2, 6, 7, 5, 3, 4 ]$  $
  [ 1, 2, 7, 4, 6, 5, 3 ]$ $  [ 1, 4, 2, 3, 6, 7, 5 ]$  $ [ 1, 6, 2, 7, 4, 3, 5 ]$ $  [ 1, 5, 2, 7, 6, 3, 4 ]$  $
  [ 1, 6, 4, 2, 3, 5,7 ]$ $  [ 1, 6, 5, 2, 3, 7, 4 ]$\\ 
  \hline 
  \cellcolor{black!20!white}{ $5$} & $3651$ & $[ 3, 4, 1, 2, 6, 5,7 ]$  $ [ 5, 2, 1, 7, 3, 4, 6 ] $&  $42$&$57$ & $[ 1, 2, 5, 3, 7, 6, 4 ]$  $ [ 1, 2, 7, 6, 4, 3, 5 ]$\\ 
  \hline 
  \cellcolor{black!20!white}{ $6$} & $2811$ & $[ 7, 2, 1, 5, 6, 4, 3 ]$  $ [ 3, 4, 2, 5, 7, 1, 6 ]$&  $21$&$166$ & $[ 1, 2, 3, 7, 6, 5, 4 ]$ $  [ 1, 2, 7, 4, 6, 5, 3 ]$  $ [ 1, 6, 2, 5, 4, 7, 3 ]$\\ 
  \hline 
   \cellcolor{black!20!white}{ $7$} & $1684$ & $[ 6, 2, 4, 3, 7, 1, 5 ]$  $ [ 1, 3, 6, 5, 7, 2, 4 ] $&  $42$&$3$ & $-$\\ 
  \hline 
   \cellcolor{black!20!white}{ $8$} & $1181$ & $[ 2, 5, 7, 3, 4, 1, 6 ] $&  $7$&$624$ & $[ 1, 2, 7, 6, 3, 5, 4 ]$ $  [ 1, 5, 7, 2, 4, 3, 6 ]$  $ [ 1, 5, 6, 3, 2, 7, 4 ]$\\ 
  \hline
   \cellcolor{black!20!white}{ $9$} & $686$ & $[ 2, 5, 7, 4, 1, 3, 6 ] $&  $3$&$1418$ & $[ 1, 7, 2, 4, 6, 5, 3 ]$ $  [ 1, 6, 3, 4, 7, 5, 2 ]$  $ [ 4, 6, 2, 1, 7, 3, 5 ]$ $  [ 3, 6, 5, 1, 2, 7, 4 ]$\\ 
  \hline  
   \cellcolor{black!20!white}{ $10$} & $475$ & $[ 2, 4, 7, 5, 3, 6, 1 ] $&  $6$&$92$ & $[ 3, 6, 4, 2, 5, 1,7 ]$\\ 
  \hline 
    \cellcolor{black!20!white}{ $11$} & $219$ & $[ 6, 5, 3, 7, 2, 1, 4 ]  $&  $2$&$1400$ & $[ 1, 7, 3, 5, 6, 4, 2 ]$ $  [ 5, 4, 3, 2, 1, 7, 6 ]$  $ [ 7, 2, 1, 4, 6, 5, 3 ]$\\ 
  \hline
    \cellcolor{black!20!white}{ $12$} & $163$ & $[ 2, 4, 7, 6, 3, 5, 1 ] $&  $7$&$40$ & $-$\\ 
  \hline
    \cellcolor{black!20!white}{ $13$} & $83$ & $[ 1, 7, 6, 4, 5, 3, 2 ] $&  $2$&$198$ & $ [ 6, 2, 4, 5, 7, 3, 1 ]$\\ 
  \hline  
   \cellcolor{black!20!white}{ $14$} & $66$ & $ [ 6, 5, 3, 4, 2, 1 ,7]$&  $2$&$266$ & $[ 7, 5, 1, 4, 3, 6, 2 ]$\\ 
  \hline    
\end{tabular}
\caption{\small{New $(7,d)-$codes  and some details of software checking.}}\label{detail7}
\end{center}
\end{table}

\begin{table}[!tbp]
\begin{center}
\begin{tabular}{ | m{.2cm} | m{.9cm}| m{7.5em} |m{.4cm}  |m{.55cm} |m{7.5em} |} 

  \hline
  \cellcolor{black!20!white}{$d$} & \cellcolor{black!20!white}{$|C_7^d|$} & \cellcolor{black!20!white}{Generators of $H$}&  \cellcolor{black!20!white}{$|H|$}&\cellcolor{black!20!white}{$\lambda_H$} & \cellcolor{black!20!white}{Elements of $S_H$}\\ 
  \hline
 \cellcolor{black!20!white}{ $3$} & $105236$ & $ [ 4, 2, 5, 1, 3, 8, 7, 6 ]$  $ [ 8, 6, 3, 4, 1, 7, 2, 5 ] $&  $336$&$120$ & $[ 1, 2, 3, 4, 5, 8, 7, 6 ]$ $ [ 1, 2, 3, 4, 6, 8, 5, 7 ]$  $ [ 1, 2, 3, 7, 4, 5, 6 ,8]$ $ [ 1, 2, 3, 8, 4, 7, 5, 6 ]$  $
  [ 1, 2, 3, 7, 4, 8, 6, 5 ]$ $ [ 1, 2, 3, 6, 5, 4, 8, 7 ]$  $ [ 1, 2, 3, 8, 5, 4, 6, 7 ]$ $ [ 1, 2, 3, 8, 6, 4, 5, 7 ]$  $
  [ 1, 2, 3, 7, 5, 8, 4, 6 ]$ $ [ 1, 2, 3, 7, 8, 6, 5, 4 ] $\\ 
  \hline 
  \cellcolor{black!20!white}{ $4$} & $89682$ & $[ 7, 1, 8, 3, 4, 2, 6, 5 ]$  $ [ 6, 5, 4, 2, 3, 8, 1, 7 ]  $&  $168$&$240$ & $[ 1, 2, 3, 4, 8, 5, 7, 6 ]$ $ [ 1, 2, 3, 4, 6, 8, 7, 5 ]$  $ [ 1, 2, 3, 7, 4, 5, 8, 6 ]$ $ [ 1, 2, 3, 5, 8, 4, 6, 7 ]$  $
  [ 1, 2, 3, 6, 5, 4, 8, 7 ]$ $ [ 1, 2, 3, 6, 7, 4, 5,8 ]$  $ [ 1, 2, 3, 8, 7, 4, 6, 5 ]$ $ [ 1, 2, 3, 5, 7, 6, 4 ,8]$  $
  [ 1, 2, 3, 8, 5, 7, 6, 4 ]$ $ [ 1, 2, 3, 6, 8, 7, 5, 4 ]$  $ [ 1, 2, 5, 4, 3, 6, 8, 7 ]$ $ [ 1, 2, 8, 5, 4, 3, 6, 7 ]$\\ 
  \hline 
  \cellcolor{black!20!white}{ $5$} & $66442$ & $ [ 7, 2, 8, 6, 5, 4, 1, 3 ]$  $ [ 6, 4, 3, 5, 2, 8, 7, 1 ] $&  $336$&$16$ & $[ 1, 2, 3, 8, 4, 7, 5, 6 ]$\\ 
  \hline 
   \cellcolor{black!20!white}{ $6$} & $54709$ & $ [ 8, 3, 4, 6, 5, 7, 1, 2 ]$  $ [ 5, 2, 4, 8, 3, 1, 6, 7 ] $&  $56$&$672$ & $ [ 1, 2, 3, 8, 4, 6, 7, 5 ]$ $ [ 1, 2, 3, 7, 6, 4, 8, 5 ]$ $ [ 1, 2, 3, 5, 8, 6, 7, 4 ]$ $ [ 1, 2, 6, 5, 3, 4, 8, 7 ]$ $
  [ 1, 2, 7, 5, 3, 6, 8, 4 ]$ $ [ 1, 2, 7, 8, 6, 3, 5, 4 ] $\\ 
  \hline 
   \cellcolor{black!20!white}{ $7$} & $37499$ & $[ 8, 5, 4, 1, 6, 3, 7, 2 ]$  $ [ 7, 2, 1, 3, 6, 8, 5, 4 ] $&  $56$&$390$ & $[ 1, 2, 7, 6, 3, 4, 5 ]$ $ [ 1, 2, 4, 6, 7, 8, 5, 3 ]$\\ 
  \hline
   \cellcolor{black!20!white}{ $8$} & $29249$ & $[ 5, 3, 6, 1, 2, 8, 7, 4 ]$  $ [ 7, 2, 6, 8, 4, 5, 3, 1 ]  $&  $56$&$390$ & $[ 1, 2, 4, 8, 5, 7, 3, 6 ] $\\ 
  \hline  
   \cellcolor{black!20!white}{ $9$} & $18352$ & $[ 4, 1, 7, 6, 8, 3, 5, 2 ]$  $ [ 5, 1, 7, 3, 2, 6, 4,8 ] $&  $48$&$12$ & $-$\\ 
  \hline 
    \cellcolor{black!20!white}{ $10$} & $13529$ & $[ 6, 1, 3, 5, 7, 2, 4,8 ]$  $ [ 8, 7, 1, 6, 3, 2, 4, 5 ] $&  $24$&$260$ & $[ 1, 2, 7, 6, 5, 3, 8, 4 ]$\\ 
  \hline
    \cellcolor{black!20!white}{ $11$} & $8135$ & $[ 5, 6, 8, 7, 1, 2, 4, 3 ]$  $[ 1, 3, 7, 8, 5, 2, 4, 6 ] $&  $12$&$212$ & $[ 2, 8, 3, 1, 6, 4, 7, 5 ]$\\ 
  \hline
    \cellcolor{black!20!white}{ $12$} & $6163$ & $[ 7, 8, 5, 6, 3, 4, 1, 2 ]$  $[ 4, 8, 5, 2, 7, 3, 6, 1 ] $&  $24$&$12$ & $-$\\ 
  \hline  
   \cellcolor{black!20!white}{ $13$} & $3169$ & $ [ 4, 6, 1, 5, 8, 3, 7, 2 ]$&  $7$&$708$ & $ [ 3, 7, 4, 6, 2, 5, 1,8 ] $\\ 
  \hline    
   \cellcolor{black!20!white}{ $14$} & $2324$ & $ [ 4, 6, 1, 5, 8, 3, 7, 2 ]$&  $7$&$708$ & $ [ 3, 7, 4, 6, 2, 5, 1,8 ] $\\ 
  \hline
   \cellcolor{black!20!white}{ $15$} & $810$ & $ [ 4, 6, 7, 8, 1, 3, 5, 2 ]$&  $8$&$168$ & $ -$\\ 
  \hline 
       \cellcolor{black!20!white}{ $16$} & $607$ & $ [ 4, 7, 6, 8, 1, 2, 3, 5 ]$  $ [ 7, 4, 5, 3, 2, 1, 8, 6 ] $&  $8$&$96$ & $ -$\\ 
  \hline 
   \cellcolor{black!20!white}{ $17$} & $252$ & $ [ 7, 3, 8, 2, 6, 1, 5, 4 ]$ &  $4$&$112$ & $ -$\\ 
  \hline
   \cellcolor{black!20!white}{ $18$} & $189$ & $ [ 7, 3, 8, 2, 6, 1, 5, 4 ]$ &  $4$&$112$ & $ -$\\ 
  \hline    
\end{tabular}
\caption{\small{New $(8,d)-$codes  and some details of software checking.}}\label{detail8}
\end{center}
\end{table}

\end{document}